\documentclass[prl,aps,floatfix,showpacs,twocolumn,reprint,superscriptaddress]{revtex4-1}

\usepackage{amssymb,amsfonts,amsmath}
\usepackage{graphics}
\usepackage[dvips]{graphicx}
\usepackage{color}
\usepackage{subfigure}
\usepackage{mathtools}
\usepackage[normalem]{ulem}
\usepackage{dcolumn}
\usepackage{multirow}
\usepackage{booktabs}
\usepackage{bm}
\usepackage[linktocpage,colorlinks=true,linkcolor=blue,citecolor=blue]{hyperref}

\begin{document}

\title{Autonomous Motility of Active Filaments due to Spontaneous Flow-Symmetry Breaking}

\author{Gayathri Jayaraman}
\affiliation{The Institute of Mathematical Sciences, CIT Campus, Chennai 600113, India}
\author{Sanoop Ramachandran}
\affiliation{Department of Physics, Indian Institute of Technology Madras, Chennai 600036, India}
\author{Somdeb Ghose}
\affiliation{The Institute of Mathematical Sciences, CIT Campus, Chennai 600113, India}
\author{Abhrajit Laskar}
\affiliation{The Institute of Mathematical Sciences, CIT Campus, Chennai 600113, India}
\author{M. Saad Bhamla}
\affiliation{Department of Physics, Indian Institute of Technology Madras, Chennai 600036, India}
\author{P. B. Sunil Kumar}
\affiliation{Department of Physics, Indian Institute of Technology Madras, Chennai 600036, India}
\author{R. Adhikari}
\affiliation{The Institute of Mathematical Sciences, CIT Campus, Chennai 600113, India}

\date{\today}

\begin{abstract}
We simulate the nonlocal Stokesian hydrodynamics of an elastic filament which is active due to a permanent distribution of stresslets along its contour. A bending instability of an initially straight filament spontaneously breaks flow symmetry and leads to autonomous filament motion which, depending on conformational symmetry can be translational or rotational. At high ratios of activity to elasticity, the linear instability develops into nonlinear fluctuating states with large amplitude deformations. The dynamics of these states can be qualitatively understood as a superposition of translational and rotational motion associated with filament conformational modes of opposite symmetry. Our results can be tested in molecular-motor filament mixtures, synthetic chains of autocatalytic particles or other linearly connected systems where chemical energy is converted to mechanical energy in a fluid environment. 
\end{abstract}


\maketitle

Components which convert chemical energy to mechanical energy internally are ubiquitous in biology. Common examples where this conversion leads to autonomous propulsion are molecular motors (at the subcellular level) and bacteria (at the cellular level)  \cite{nedelec1997,*camazine2003}. Recently, biomimetic elements which convert chemical energy into translational \cite{paxton2004,*vicario2005} or rotational  \cite{ozin2005,*catchmark2005} motion have been realized in the laboratory. While the detailed mechanisms leading to autonomous propulsion in these biological and soft matter systems show a wonderful variety \cite{gibbs2009}, their collective behavior tends to be universal and can be understood by appealing to symmetries and conservation laws \cite{simha2002a}. This realization has led to many studies of the collective properties of suspensions of hydrodynamically interacting autonomously motile particles \cite{cisneros2007,*saintillan2008a}.

There are ample instances in biology, however, where the conversion of chemical to mechanical energy is not confined to a particle-like element but is, instead, distributed over a line-like element. Such a situation arises, for example, in a microtubule with a row of molecular motors converting energy while walking on it. The mechanical energy thus obtained not only produces motion of the motors but also generates reaction forces on the microtubule, which can deform elastically in response. Hydrodynamic interactions between the motors and between segments of the microtubule must be taken into account since both are surrounded by a fluid. This combination of elasticity, autonomous motility through energy conversion and hydrodynamics is found in biomimetic contexts as well. A recent example is provided by mixtures of motors which crosslink and walk on polymer bundles. A remarkable cilia-like beating phenomenon is observed in these systems \cite{sanchez2011}. A polymer in which the monomeric units are autocatalytic nanorods provides a nonbiological example of energy conversion on linear elastic elements. Though such elements are yet to be realized in the laboratory, active elements coupled to passive components through covalent bonds have been synthesized \cite{vicario2005} and may lead to new kinds of nanomachines \cite{ozin2005}.

Motivated by these biological and biomimetic examples, we study, in this Letter, a semiflexible elastic filament immersed in a viscous fluid with energy converting ``active'' elements distributed along its length. We present an equation of motion for the filament that incorporates the effects of nonlinear elastic deformation, active processes and nonlocal Stokesian hydrodynamic interactions. We use the lattice Boltzmann (LB) method to numerically solve the active filament equation of motion.  Our simulations show that a straight active filament is linearly unstable to transverse perturbations. Depending on the symmetry of the perturbation, the hydrodynamic flows produce translational or rotational motion of the filament. Conformational symmetry provides a qualitative way of understanding the dynamics of the nonlinear fluctuating states that develop at high ratios of activity to elasticity. We describe these results and our model in detail below. 

%
\emph{Model:} Our model for the active filament consists of $N$ beads, with coordinates ${\bf r}_{n}$, interacting through a potential given by
%
\begin{eqnarray} \label{eq:Potential}
U({\bf r}_1, \ldots, {\bf r}_N) &=& \sum_{m=1}^{N-1} U_{\rm S} ({\bf b}_{m})+ \sum_{m=1}^{N-2} U_{\rm B}({\bf b}_{m}, {\bf b}_{m+1})\nonumber\\
&+& {\frac{1}{2}}\sum_{m, n = 1}^{N}U_{\rm LJ}({\bf r}_n - {\bf r}_m).
\end{eqnarray}
%
The two-body harmonic spring potential $U_{\rm S}({\bf b}_m) = \frac{1}{2} k (  b_m  - b_0)^2 $ penalizes departures of $b_m$, the modulus of the bond vector $ {\bf b}_m= | {\bf r}_m - {\bf r}_{m+1} |$, from its equilibrium value of $b_0$. The three-body bending potential $U_{\rm B}({\bf b}_{m}, {\bf b}_{m+1}) =\bar\kappa ( 1 -  \cos \phi_m)$ penalizes departures of the angle $\phi_m$ between consecutive bond vectors from its equilibrium value of zero. The rigidity parameter $\bar\kappa$ is related to the bending rigidity as $\kappa = b_0\bar\kappa$. The repulsive Lennard-Jones potential $U_{\rm LJ}$ vanishes if the distance between beads $r_{mn}=| {\bf r}_m - {\bf r}_{n} |$ exceeds $\sigma_{\rm LJ}$. The $n$-th bead experiences a force ${\bf f}_n = - \partial U / \partial {\bf r}_n$ when the filament stretches or bends from its equilibrium position. With the above choice of potential the connected beads approximate an inextensible, semiflexible, self-avoiding filament. 

Active nonequilibrium processes, such as those that convert chemical energy to mechanical energy, are internal to the fluid and hence cannot add net momentum to it. Then, the integral of the force density on a surface enclosing the active element must vanish. This is ensured if the active force density is the divergence of a stress. Since the active processes cannot add angular momentum to the fluid, the stress must be symmetric \footnote{J.-F. Joanny, private communication}. The most dominant Stokesian singularity with these properties is the stresslet \cite{chwang1975}. There is a remaining freedom of the sign of the stresslet and its angle relative to the filament. Motivated by the tangential stresses exerted by motors walking on microtubules \cite{sanchez2011}, we choose the stresslet to be extensile and oriented along the instantaneous tangent $\hat {\bf t}_n$ to the filament,
%
\begin{equation} \label{eq:ActiveStress}
{\boldsymbol \sigma}_n = \sigma_0(\hat{\bf t}_n \hat {\bf t}_n - {\mathbb I}/d)
\end{equation}
%
where $d$ is the spatial dimension and  $\sigma_0 > 0$ sets the scale of the activity. The results of other choices of sign and orientation will be presented elsewhere. 
%
 \begin{figure}[t!]  
 \centering
 \subfigure[~]{\includegraphics[width=0.5\textwidth]{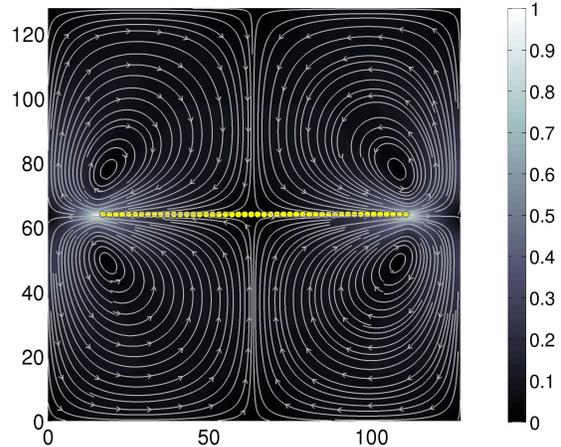} \label{fig:2dStab}} \\
 \subfigure[~]{\includegraphics[width=0.5\textwidth]{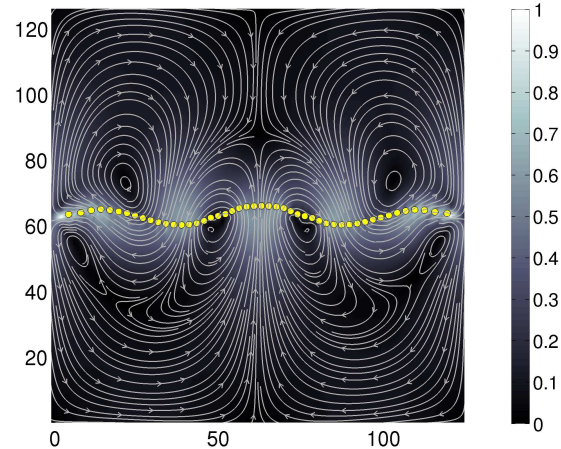} \label{fig:2dInstab}}
 \caption{(Color online) \label{fig:2dInst} 
  Spontaneous symmetry breaking due to a linear instability.
 (a) A straight filament produces a flow symmetric about both its axis and its midpoint. This flow can only produce an extensile motion of the filament. 
 (b) The straight conformation, however, is linearly unstable to transverse perturbations. Here the perturbation is symmetric about an axis passing through the filament midpoint. This leads to a flow which is no longer symmetric about the initial axis. It enhances the instability and can produce center of mass motion of the filament. 
 Bead positions are shown as yellow filled circles, while the fluid velocity is shown as streamlines with the background color indicating its magnitude.
  } 
\end{figure}

Elastic forces and active stresses produce velocities in the fluid. In the Stokesian regime, the velocity in a three-dimensional unbounded fluid at location ${\bf r}$ produced by a force ${\bf f}$ at the origin is $v_{\alpha}({\bf r}) = O_{\alpha\beta}({\bf r})f_{\beta}$ where $O_{\alpha\beta}({\bf r}) = (\delta_{\alpha\beta} + \hat r_{\alpha}\hat r_{\beta})/8\pi\eta r$ is the Oseen tensor, Cartesian directions are indicated by Greek indices, $\eta$ is the fluid shear viscosity and $\hat r_{\alpha} = r_{\alpha}/r$. Similarly, the velocity at location ${\bf r}$ produced by a stresslet $\boldsymbol \sigma$ at the origin is $v_{\alpha}({\bf r}) = D_{\alpha\beta\gamma}({\bf r})\sigma_{\beta\gamma}$ where $D_{\alpha\beta\gamma}({\bf r}) = (-\hat r_{\alpha}\delta_{\beta\gamma} + 3\hat r_{\alpha}\hat r_{\beta}\hat r_{\gamma} )/8\pi\eta r^2$ \cite{pozrikidis1992}. In the presence of rigid or periodic boundaries the tensors $\boldsymbol O$ and $\boldsymbol D$ must be replaced by the appropriate Green's functions of Stokes flow that vanish at the boundaries or have the periodicity of the domain \cite{pozrikidis1992}. Similarly, two-dimensional Green's functions must be used when studying the motion of filaments in planar flows \cite{chwang1975}.
%
 \begin{figure*}[t!]
  \centering
  \subfigure[~]{\includegraphics[width=0.40\textwidth]{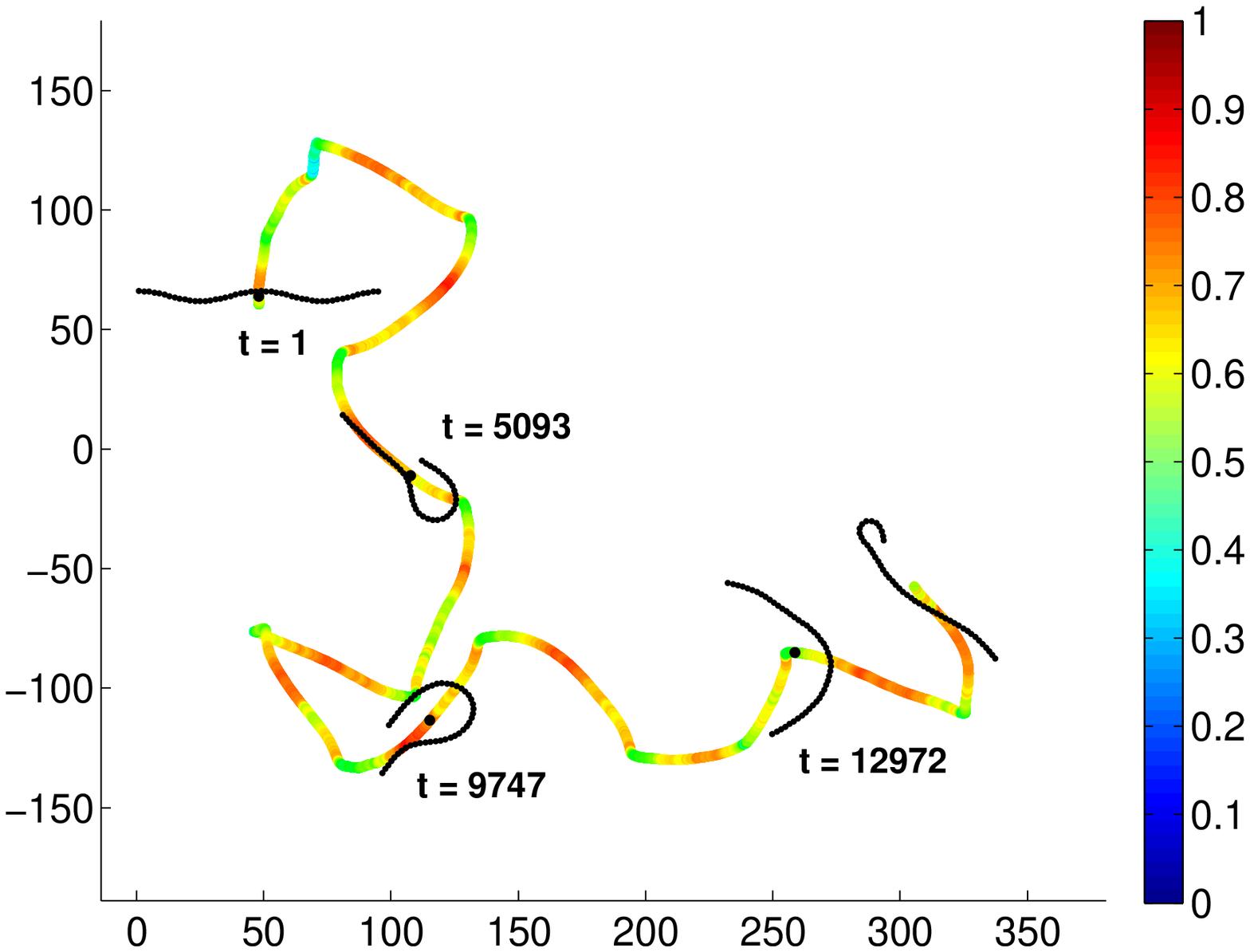} \label{fig:rNtEven}} 
  \subfigure[~]{\includegraphics[width=0.40\textwidth]{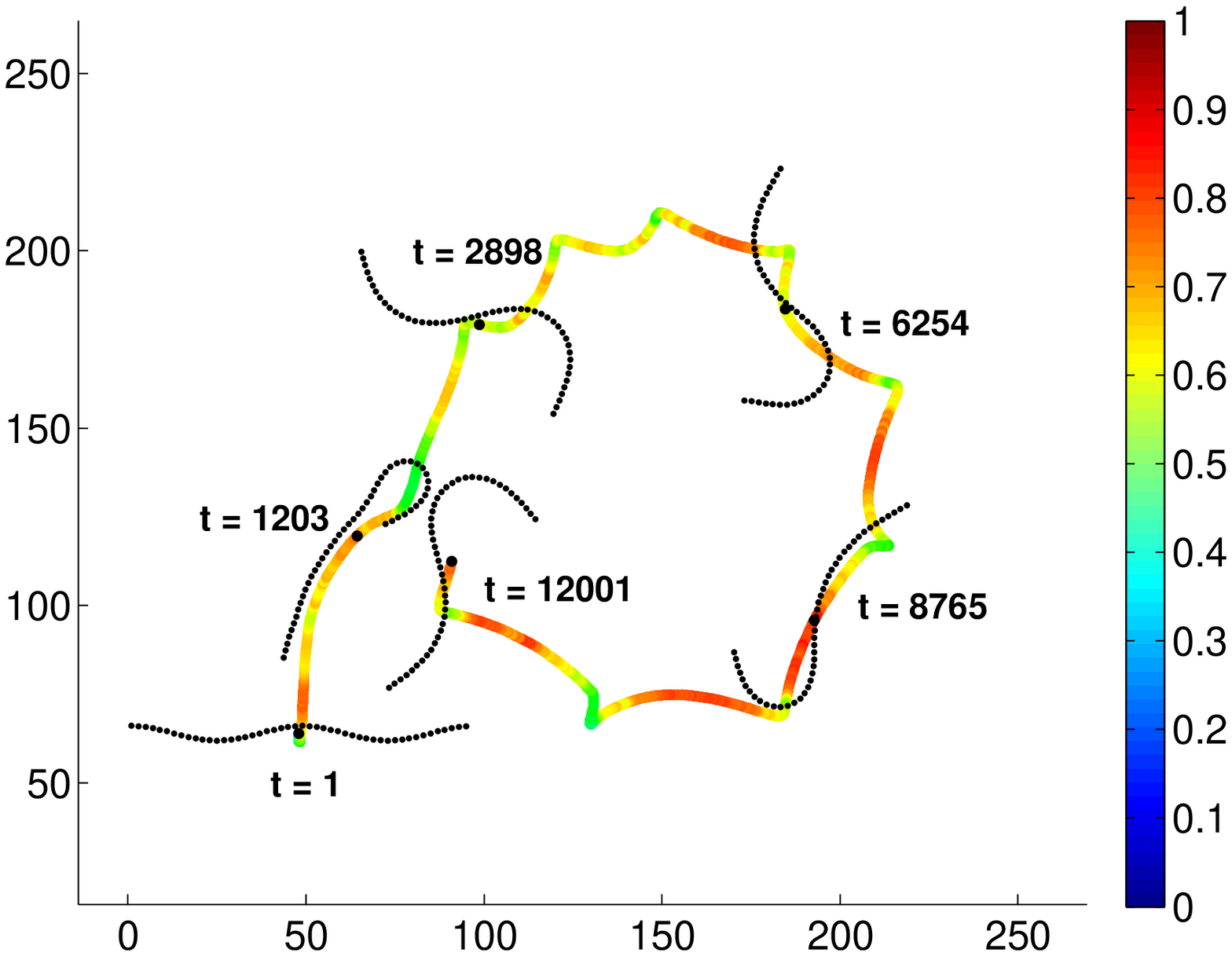} \label{fig:rNtOdd}}  \\ 
  \subfigure[~]{\includegraphics[width=0.40\textwidth]{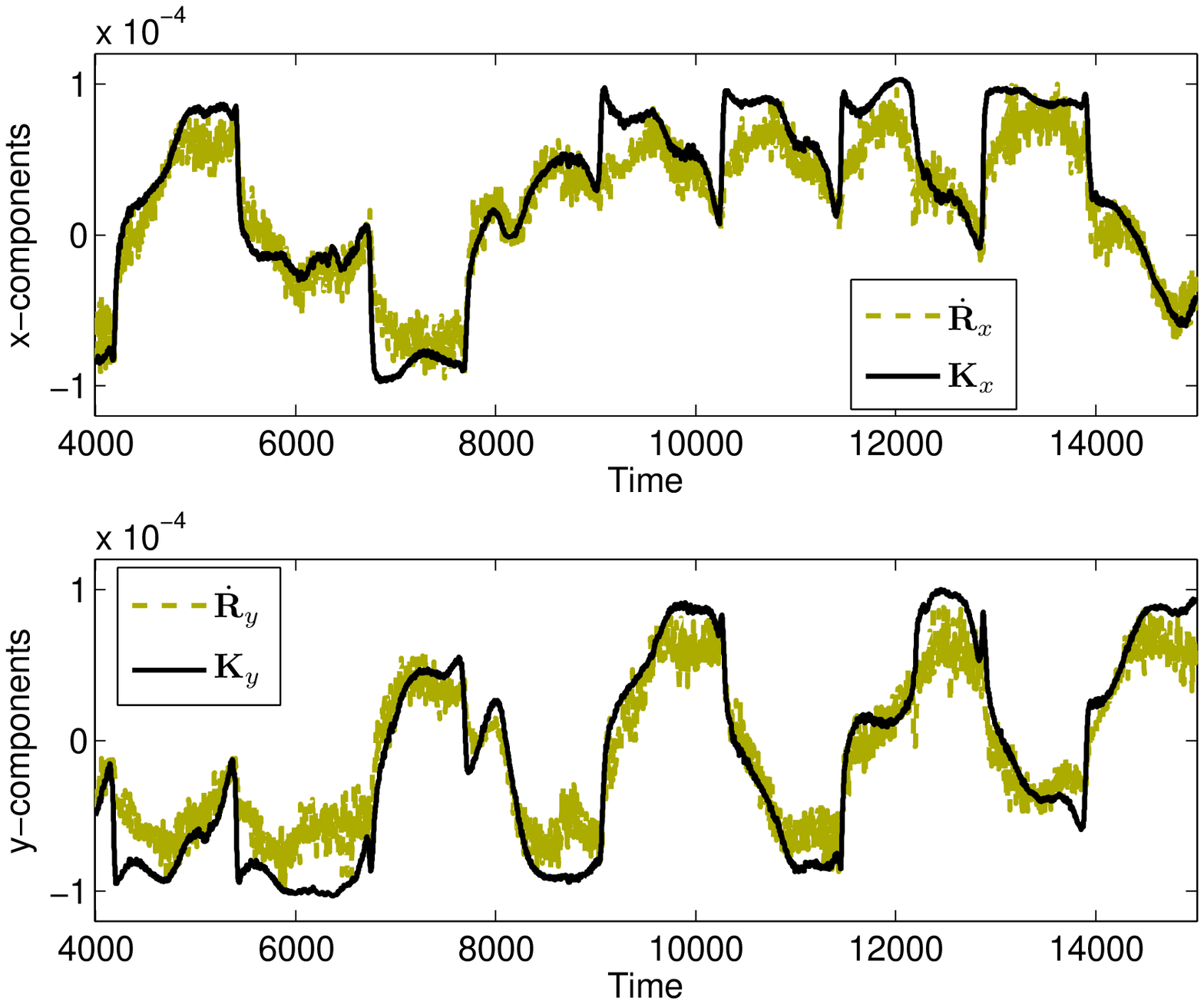} \label{fig:vCMn-A250}}
  \subfigure[~]{\includegraphics[width=0.40\textwidth]{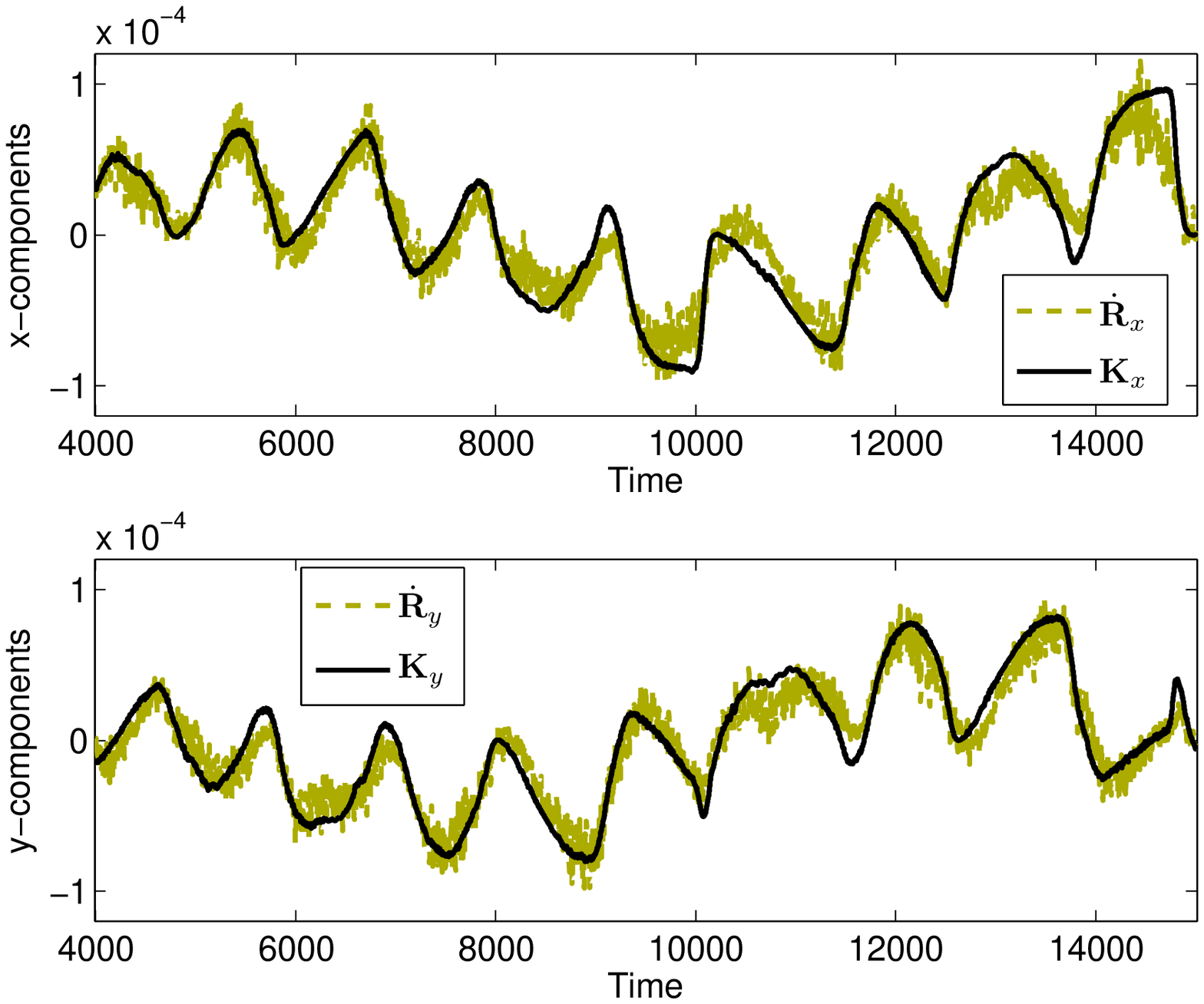} \label{fig:vCMn-A150}}
  \caption{(Color) \label{fig:Results}
  Filament motion for high activity. 
  (a) For an \emph{even} initial conformation with $\mathcal{A} = 250.7$, the motion is predominantly translational. 
  (b) For an identical initial conformation with $\mathcal{A} = 150.4$, the motion is translational as well as rotational due to the spontaneous appearance of conformations with \emph{odd} symmetry. 
  In both (a) and (b) the color of the trace corresponds to the center of mass speed $|\dot{\bf R}|$ normalized by its maximum value. 
  (c) and (d) Time traces of the $x$ (top) and $y$ components (bottom) of $\dot{\bf R}$ (olive dashed line) and ${\bf K} = - \left( \sigma_0/4\pi\eta b_0 \right) \langle \varkappa \widehat{\bf n} \rangle$ (black solid line), where $\langle \varkappa \widehat{\bf n} \rangle$ is the average curvature, corresponding to simulations in (a) and (b) respectively. The velocity and the curvature remain strongly correlated for various filament conformational modes and activity numbers. 
  Times are in $10^3$ LB steps.
  }
 \end{figure*}
%
The velocity of the $n$-th bead is obtained by summing the force and activity contributions from all  beads, including itself, to the fluid velocity at its location. An isolated spherical bead with a force ${\bf f}$ acquires a velocity $\mu \, {\bf f}$ where $\mu$ is its mobility. By symmetry, an isolated spherical bead with a stresslet $\boldsymbol \sigma$ cannot acquire a velocity. This gives the following equation of motion for the active filament:
%
\begin{equation}\label{eq:actPolEoM}
\dot {\bf r}_n = \sum_{m=1}^N\left [ {\mathbf O}({\bf r}_n - {\bf r}_m)\cdot {\bf f}_m+ {\mathbf D}({\bf r}_n - {\bf r}_m)\cdot{\boldsymbol \sigma}_m \right]
\end{equation}
%
where, for $m=n$, $ O_{\alpha\beta} = \mu\delta_{\alpha\beta}$ and $D_{\alpha\beta\gamma} = 0$. Equations (\ref{eq:Potential}), (\ref{eq:ActiveStress}) and (\ref{eq:actPolEoM}) represent our model for the nonlocal Stokesian hydrodynamics of an active elastic filament. In the absence of bending rigidity and activity, our model reduces to Zimm dynamics of a polymer in a good solvent \cite{doi1988}. 
 
The ratio of the stresslet and Stokeslet terms in the equation of motion is a dimensionless measure of activity. Estimating the curvature elastic force as $\kappa/ L^2$, where $L = (N-1) b_0$ is the length of the filament, yields the ``activity number" $\mathcal{A} = L\sigma_0/\kappa$. The rates of active and elastic relaxation are  $\Gamma_{\sigma} = \sigma_0/\eta L^d$ and $\Gamma_{\kappa} = \kappa/\eta L^{d+1}$ respectively. Since $\mathcal{A} = \Gamma_{\sigma}/\Gamma_{\kappa}$, the activity number also measures the ratio of time scales associated with active and elastic relaxation. As $\mathcal{A}\rightarrow 0$ the active time scale diverges and  conformational changes occur only due to elastic forces. As $\mathcal{A}\rightarrow\infty$ conformational changes due to activity are much more rapid than those due to elasticity.

\emph{Method:} 
We use the lattice Boltzmann method \cite{benzi1992,*succi2001,*aidun2010} to obtain the incident flow in Eq.\ (\ref{eq:actPolEoM}), corresponding to terms with $m\neq n$, and then integrate the equations using the forward Euler method. The method of obtaining the incident flow is described in detail in \cite{nash2008,*nash2010} and in \cite{SI}, and is related to methods used in \cite{ahlrichs1999,*fyta2008,*pham2009}. The subtleties in using the lattice Boltzmann method to solve Eq.\ (\ref{eq:actPolEoM}) are described in detail in \cite{SI}. We use lattice units in which both spatial and temporal discretization scales are unity. We choose $b_0 = 2$ and $k$ such that there is less than $1\%$ variation in contour length. We choose $\bar\kappa$ in the range $0.0$ to $0.5$, $\sigma_0$ in the range $0$ to $0.05$, and $N$ in the range $16$ to $96$. The initial filament conformation is a superposition of small amplitude transverse sinusoidal deformations of wavelengths a few integer multiples of $L$. The integration is carried out for several million time steps. Our results, unless otherwise stated, are for periodic planar lattices of size $128$.


\emph{Results:} 
We summarize our results in Figs.\ (\ref{fig:2dInst}) and (\ref{fig:Results}) and the movies in \cite{SI}. Our simulations find a linear instability of an initially straight filament. On dimensional grounds, this instability occurs only when $L > l_{\mathcal{A}}\sim \kappa /\sigma_0$,  where numerical prefactors can be obtained from a linear stability analysis of Eq.\ (\ref{eq:actPolEoM}). In contrast, the elastic Euler instability of a filament under force $F$ occurs when  $L > l_{E}\sim \sqrt{\kappa / F}$. A linear instability of passive filaments in an active medium without nonlocal hydrodynamics was reported in \cite{kikuchi2009}, while bow-shaped conformation for filaments driven by external forces were reported in \cite{xu1994,*lagomarsino2005}.

We show the nature of the linear instability, as $\mathcal{A} \rightarrow \infty$ and $k\rightarrow 0$, in Fig.\ (\ref{fig:2dInst}) and its accompanying movie \cite{SI}. The flow produced by a straight filament is symmetric about the filament center and the filament axis, as shown in Fig.\ \ref{fig:2dStab}. It can only produce an extensile motion in the filament which is transient, since it is balanced by the stretching elasticity. A spontaneous transverse perturbation breaks the flow symmetry about the initial axis, enhancing the perturbation, as shown in Fig.\ \ref{fig:2dInstab}. The flow now develops an uncompensated directional component in which the filament can translate. 
Since the hydrodynamic drag on the filament is greater at its ends \cite{keller1976}, a balance between elastic deformation, active propulsion and drag ensues and the filament propels steadily in a deformed bow-shaped conformation \cite{SI}.

In Fig.\ (\ref{fig:2dInst}), the initial perturbation is symmetric about the filament midpoint. We call this an \emph{even} mode.  Initial perturbations which are antisymmetric about the midpoint are also linearly unstable. However, these \emph{odd} modes produce flows of a completely
different nature than the \emph{even} modes. Instead of an uncompensated linear component, the flow develops a vorticity centered on the filament midpoint which results in pure rotational motion \cite{SI}. A generic initial perturbation is a superposition of both \emph{even} and \emph{odd} modes
and, thus, both rotates and translates. At low $\mathcal{A}$, there is little coupling between the conformational modes, due to which the filament has steady motion. However, with increasing $\mathcal{A}$ greater elastic deformations are needed to balance the activity, due to which the filament develops nonlinear fluctuating states with large-amplitude deformations, as shown in Fig.\ (\ref{fig:Results}) and the accompanying movies \cite{SI}. Conformational modes are now coupled, and modes absent from the initial perturbation can spontaneously appear. With a predominance of \emph{even} modes, the motion is primarily translational as seen in Fig.\ \ref{fig:rNtEven}, but when a spontaneously generated \emph{odd} mode appears the motion is both translational and rotational as seen in Fig.\ \ref{fig:rNtOdd}. 
In cubic flows, we see similar linear instabilities and nonlinear fluctuating states \cite{SI}.
It is surprising that such complex ``animate'' behavior can be generated by Eq.\ (\ref{eq:actPolEoM}). Remarkably, its qualitative aspects can be understood from basic symmetry arguments relating conformations to the flows they produce. The role of symmetry in the motility of active droplets has been studied recently in \cite{tjhung2012}.

In the free-draining approximation, it is possible to relate the center of mass velocity $\dot{\bf R}$ to the mean curvature of the filament \cite{SI},
\begin{equation}
 \dot {\bf R} =  - \frac{\sigma_0}{4\pi\eta b_0}\langle \varkappa \widehat{\bf n} \rangle
\end{equation}
where $\varkappa$ is the local curvature and $\widehat{\bf n}$ is the local unit normal vector.
In Figs.\ \ref{fig:vCMn-A250} and \ref{fig:vCMn-A150} we plot components of this equation corresponding to simulations in Figs.\ \ref{fig:rNtEven} and \ref{fig:rNtOdd} respectively. The agreement is good in both cases. Thus local hydrodynamics provides a good approximation for the translational velocity but non-local hydrodynamics is required to fully explain the conformational fluctuations. The interplay between nonlocal hydrodynamics and semiflexibility is necessary for the rotational motion of the filament, as has been noted earlier in a different context \cite{lagomarsino2005}.

For a microtubule of size $L \sim 30\mu m$, $\kappa \sim 50 pN\mu m^2$ with about $ 200$ motors per micron exerting approximately $6pN$ of force,  we obtain $\mathcal{A} \sim 60$. 
These parameters provide a translational speed of $\sim 1 mm s^{-1}$ for a semicircular shape in water.
The activity can be manipulated in motor-polymer bundle systems or in polymers of autonomously motile nanorods over a large range of $\mathcal{A}$. These systems would be the best candidates for a verification of our results.


\emph{Discussion and conclusion:}
Our model has several important variations. We argued that active processes cannot add linear or angular momentum to the fluid and, so, must be represented by Stokesian singularities with those properties. This ruled out the Stokeslet and the rotlet but allowed for higher singularities, of which the stresslet, being the most dominant, was retained. The stresslet, with a $\boldsymbol C_{\infty}$ axis, is not forbidden by symmetry as a representation of a polar active element. If it is forbidden for non-symmetry reasons, we must use the potential dipole $\boldsymbol d$ \cite{pozrikidis1992}, the leading singularity with polar symmetry, whose velocity field is $v_{\alpha}({\bf r}) = G_{\alpha\beta}({\bf r})d_{\beta}$,  $G_{\alpha\beta} =  (-\delta_{\alpha\beta}+ 3\hat r_{\alpha}\hat r_{\beta})/8\pi\eta r^3$, in Eq.\ (\ref{eq:actPolEoM}). The axis of the stresslet or the potential dipole can be oriented normally or obliquely to the local tangent of the filament and the stresslet can also be contractile, $\sigma_0 < 0$. The precise nature of the nonlinear steady states obtained from these various combination will be reported in future work.  A generic equation of motion encompassing these specific cases is provided in \cite{SI}.

Semiflexibility is crucially important in obtaining the results above. A rigid rod ($\kappa = \infty$, $\mathcal{A}=0$) is immune to the active instability. Since the uniaxial axes of the stresslets and the rod are aligned, it cannot, by symmetry, acquire any translational or rotational motion. It is only by the breaking of this symmetry, possible when $\mathcal{A} \neq 0$, that the filament is able to acquire motion. 

We have presented a model for an autonomously motile semiflexible filament which takes into account nonlocal hydrodynamic interactions. Our model opens up the possibility of studying the nonequilibrium dynamics of active filaments, for example the cilia-like beating of motor-polymer bundles \cite{sanchez2011} as well as the collective properties of networks of active filaments, such as the cytoskeleton.


Financial support from PRISM, Department of Atomic Energy, Government of India (AL, SG and RA) and computing resources through HPCE, IIT Madras and Annapurna, IMSc are gratefully acknowledged. The authors thank M.E. Cates, D. Frenkel and I. Pagonabarraga for useful suggestions. RA thanks the Cambridge-Hamied Visiting Lecture Scheme for supporting a visit to Cambridge University where part of this work was conceptualized.

\bigskip
\noindent \textbf{Supplemental material:}

\begin{figure}[!ht]  
 \centering
 \subfigure[~]{\includegraphics[width=0.435\textwidth]{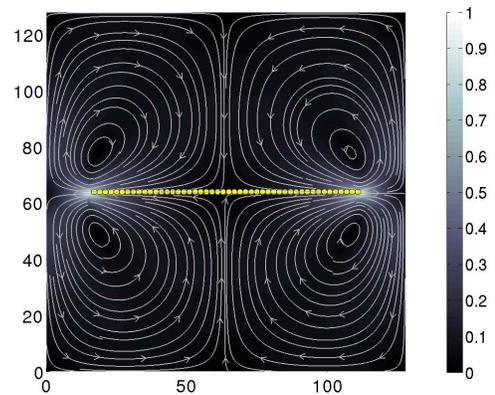} \label{fig:SI:1a_LB}} 
 \subfigure[~]{\includegraphics[width=0.435\textwidth]{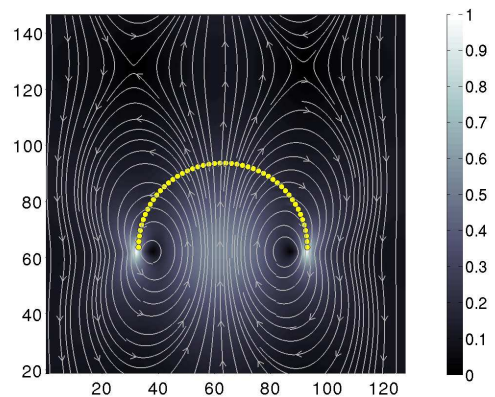} \label{fig:SI:1b_LB}}
 \subfigure[~]{\includegraphics[width=0.435\textwidth]{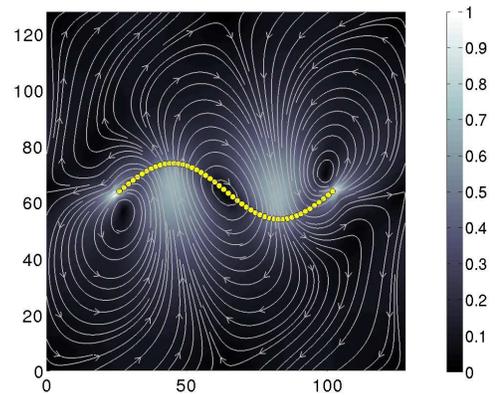} \label{fig:SI:1c_LB}} 
 \caption{\label{fig:SI:SD-LB-comparison}FIG.\ S1. Effect of the spontaneous breaking of symmetry on fluid flows, which for the motile (b) \emph{even} and (c) \emph{odd} symmetry broken conformations can be clearly distinguished from that of (a), where symmetry is preserved and no autonomous motion is possible. Bead positions are shown as yellow filled circles, while the fluid velocity is shown as streamlines with the background color indicating its magnitude. 
 } 
\end{figure}

\emph{Symmetries and flows} : Spontaneous symmetry breaking of the active filament generates net flows in which the filament can translate and/or rotate autonomously. In the absence of symmetry breaking, as seen in Fig.\ S\ref{fig:SI:1a_LB}, the net flow on the filament cancels and no motion is produced. Symmetry broken conformations can be distinguished as \emph{even} or \emph{odd} if they are symmetric or anti-symmetric, respectively, about an axis passing through the filament midpoint. In Fig.\ S\ref{fig:SI:1b_LB}, an \emph{even} conformation produces a net translatory flow, with no rotational components. In Fig.\ S\ref{fig:SI:1c_LB}, an \emph{odd} conformation produces a net rotatory flow, with no translational components. The most general symmetry-broken conformation of the filament is a linear combination of \emph{even} and \emph{odd} conformations, allowing it to translate as well as rotate.

\emph{Lattice Boltzmann method of solution} : The lattice Boltzmann method (LBM) is an efficient way of solving the Navier-Stokes equations. Here, we use it to solve for the flow produced by the distribution of forces and stresses produced by the active filament in a periodic geometry. The advantage of this is that it avoids the complicated Ewald summation of the hydrodynamic Green's functions. In particular, the LBM is easily extended to more complicated geometries, where closed form expressions of the hydrodynamic Green's functions are not available. Specifically, we solve
\begin{align}
\label{dbe}
\partial_t f_i +{\bf c}_i\cdot\nabla f_i  + [{\bf 
F}\cdot\nabla_{\bf c}f]_i = - {\cal L}_{ij}(f_j - f_j^0)
\end{align}
where $f_i({\bf x},t)$ is the one-particle distribution function in phase space of coordinates ${\bf x}$ and velocities ${\bf c}_i$,
${\cal L}_{ij}$ is the collision matrix linearized about the local equilibrium $f^0_i$ and the repeated index $j$ is summed over. The 
force density ${\bf F}$ is a sum of all the contributions ${\bf f}_n$ obtained from the potential energy of filament deformation and the stresslets on each monomer, modelled here as a pair of point forces separated by a distance $d$. This methodology is explained in detail in \cite{nash2008}. The flow field at the location of the $n$-th bead, ${\bf v}({\bf r}_n)$, is computed by interpolation from the LBM grid points after ``self'' contributions from the monopole and dipole singularities have been subtracted. The bead is then moved forward by the equation of motion
\begin{align} \label{eq:faxenEoM}
 \dot{\bf r}_n = {\frac{{\bf f}_n}{6\pi\eta a}} + {\bf  v}({\bf r}_n)
\end{align}
In this regard, our method is related to those of D\"{u}nweg and co-workers \cite{ahlrichs1999, *pham2009} and Fyta et al \cite{fyta2008}, where the LBM is used to solve for the fluid flow around a polymer. The main distinction is that those authors retain inertia for the beads while we work in the overdamped limit. 

To ensure that the LBM produces the Stokesian limit and that the velocity obtained above is identical to that produced by a direct summation of singularities, as in Eq.\ (3) of the main text, several conditions need to be ensured. First, the Reynolds number at the scale of the filament must be small, so that the nonlinear advection term in the Navier-Stokes equations is negligible. In our simulations, this is typically kept to about $10^{-2}$. Second, momentum diffusion across the filament must be rapid compared to the time scale of filament motion, so that there are no hydrodynamic retardation effects. In other words, the momentum diffusion time $L^2/\eta$ must be shorter than times of elastic and active relaxation. In our simulations, these ratios vary from $10^{-1}$ to $10^{-3}$. We always ensure that the LBM operates in the hydrodynamic limit of small Knudsen number by choosing LBM relaxation times (related to the shear and bulk viscosities) to be sufficiently below unity.


\emph{Free-draining approximation of filament velocity} : Consider the space curve, ${\bf r}  = {\bf r}(s)$. Taylor expanding about a point $s$ gives, 
\begin{equation}
{\bf r}(s^{\prime}) = {\bf r}(s) + \Delta s\partial_s{\bf r} + \frac{1}{2}\Delta s^2\partial_s^2{\bf r} + O(\Delta s^3)
\end{equation}
Using the Frenet-Serret relations between the tangent ${\bf t} = \partial_s{\bf r}$ and the normal ${\bf n}$ we can recast the above expansion as
\begin{equation}
{\bf r}(s^{\prime}) - {\bf r}(s) =  \Delta s {\bf t} + \frac{1}{2}\Delta s^2\varkappa {\bf n} + O(\Delta s^3)
\end{equation}
where the curvature $\varkappa$ and torsion $\tau$ are evaluated at $s$. If the curve is discretised by $N$ points, each separated by a distance $b_0$, and inextensibility is enforced so that $s$ remains a material parameter, we can write for the pairs of points ${\bf r}_{n-1}$,  ${\bf r}_{n}$ and  ${\bf r}_n$,  ${\bf r}_{n+1}$ 
\begin{align}
& {\bf r}_{n} - {\bf r}_{n-1} =  b_0 {\bf t}_n - \frac{1}{2} b_0^2 \varkappa {\bf n}_n \\
& {\bf r}_{n} - {\bf r}_{n+1} =  - b_0 {\bf t}_n - \frac{1}{2} b_0^2\varkappa {\bf n}_n 
\end{align}
The flows induced at ${\bf r}_n$ by stresslets at ${\bf r}_{n-1}$ and ${\bf r}_{n+1}$,  oriented along the local tangents, are  $(\sigma_0/4\pi\eta b_0^3)({\bf r}_{n} - {\bf r}_{n-1})$ and  $(\sigma_0/4\pi\eta b_0^3)({\bf r}_{n} - {\bf r}_{n+1})$ respectively. Utilising the above relations, the total flow at ${\bf r}_n$ due to neighbouring stresslets is
\begin{equation}
{\bf \dot r}_n = - \frac{\sigma_0}{4\pi\eta b_0} \varkappa {\bf n}_n
\end{equation}
In the nearest-neighbour approximation, the active velocity is proportional only to the curvature and is normal to the curve.In the free-draining approximation, the equation of motion for an active filament then is
\begin{equation}
\dot{\bf r}_n = \frac{{\bf f}_n}{6\pi\eta b_0} - \frac{\sigma_0}{4\pi\eta b_0}\varkappa {\bf n}_n
\end{equation}
This can be thought of as the ``Rouse'' limit of Eq.\ (3) of the main text. The center of mass velocity can be obtained by summing all the bead velocities, while recalling that the sum of all forces is zero, to get
\begin{equation}
\dot {\bf R} =  - \frac{\sigma_0}{4\pi\eta b_0}\langle \varkappa {\bf n} \rangle
\end{equation}


\emph{Generic model for autonomously motile elastic filaments} : We present a generic model for autonomously motile elastic filaments which encompasses all the variations of Eq.\ (3), discussed in the main text. We include the potential dipole as a possible singularity that is polar. This is subdominant to the stresslet, but is the most important singularity if the stresslet vanishes due to non-symmetry reasons. In addition, we include an externally imposed shear flow characterized by the shear rate tensor ${\bf E}$. We allow for any orientation of the stresslet axis ${\bf \hat p}_n$ and the potential dipole axis ${\bf \hat d}_n$ relative to the local tangent of the filament ${\bf \hat t}_n$. Thus, we have two new parameters in the model, $\theta_{\sigma} = {\bf \hat p}_n\cdot{\bf \hat t}_n$ and $\theta_{d} = {\bf \hat d}_n\cdot{\bf \hat t}_n$, the preferred angle that the stresslet and the potential dipole make with respect to the tangent. These angles can be made to vary along the filament, or may fluctuate in response to thermal noise. Finally, we include an external force ${\bf g}_n$ which may be due to externally imposed fields like gravity or electricity. Such fields are required when studying the driven motion, for example sedimentation under gravity, of active filaments. The generic equation of motion then,  is
\begin{widetext}
\begin{align} \label{eq:actPolEoMgeneric}
\dot {\bf r}_n = \sum_{m=1}^N\left [ {\mathbf O}({\bf r}_n - {\bf r}_m)\cdot {\bf f}_m+ {\mathbf D}({\bf r}_n - {\bf r}_m)\cdot{\boldsymbol \sigma}_m +   {\mathbf G}({\bf r}_n - {\bf r}_m)\cdot {\bf d}_m \right] + \mathbf{ E\cdot r}_n 
\end{align}
%
\begin{align}
{\bf f}_n =-{\frac{\partial U}{\partial {\bf r}_n}} + {\bf g}_n, \qquad {\boldsymbol \sigma}_n =\sigma_0(\hat{\bf p}_n \hat {\bf p}_n - {\mathbb I}/d), \qquad 
{\bf d}_n = d_0 {\bf  \hat d}_n.
\end{align}
%
%
\begin{align}
O_{\alpha\beta}({\bf r}) = {\frac{1}{8\pi\eta r}} (\delta_{\alpha\beta} + \hat r_{\alpha}\hat r_{\beta}), \,\,  D_{\alpha\beta\gamma}({\bf r}) ={\frac{1}{8\pi\eta r^2}} (-\hat r_{\alpha}\delta_{\beta\gamma} + 3\hat r_{\alpha}\hat r_{\beta}\hat r_{\gamma} ), \,\, G_{\alpha\beta} = {\frac{1}{8\pi\eta r^3}}(-\delta_{\alpha\beta}+ 3\hat r_{\alpha}\hat r_{\beta})
\end{align}
\end{widetext}
For an unbounded two-dimensional fluid, the tensors can be obtained from their corresponding three-dimensional expressions through the replacements $1/r \rightarrow \log r$, $1/r^{n+1} \rightarrow 1/r^n$ for $n=1, 2$ and $8\pi \rightarrow 4\pi$ \cite{chwang1975}. For periodic flows, the forms given by Hasimoto must be used \cite{hasimoto1959}. If required, the hydrodynamic interactions can be evaluated to higher orders in a multipole expansion \cite{kutteh2003} or can be formulated within the more rigorous framework of slender body theory \cite{keller1976}. 

The relaxation rates associated with elasticity, stresslet activity and potential dipole activity are  $\Gamma_{\kappa} = \kappa/\eta L^{d+1}$, $\Gamma_{\sigma} = \sigma_0/\eta L^d$ and $\Gamma_{d} = d_0/\eta L^{d+1}$. The shear rate tensor introduces at least one additional  independent relaxation rate $\Gamma_{E} = \dot\gamma$. The ratio of uniaxial and polar activities is $L\sigma_0 / d_0$, indicating that uniaxial activity dominates for long filaments. This motivates why Eq.\ (3) in the main text retains only uniaxial activity. These equations form the basis by which we can explore the nonequilibrium dynamics of active filaments under external fields or externally imposed velocity gradients.


%

\end{document}